\documentclass[twocolumn,superscriptaddress,prl]{revtex4-1}

\newcommand{\tw}{t_\mathrm{w}}

\newcommand{\xim}{\xi_\mathrm{mic}}
\newcommand{\xiM}{\xi_\mathrm{mac}}
\usepackage[T1]{fontenc}
\usepackage{graphicx}
\usepackage{amsmath}
\usepackage{amssymb}
\usepackage{color}
\usepackage{hyperref}
\makeatletter
\newcommand*{\balancecolsandclearpage}{%
  \close@column@grid
  \cleardoublepage
  \twocolumngrid
}
\makeatother
\graphicspath{{./figures/}{./}}

\begin{document}

\title{Matching microscopic and macroscopic responses in glasses}

\author{M.~Baity-Jesi}\affiliation{Institut de Physique Th\'eorique, Universit\'e Paris Saclay, CEA, CNRS, F-91191 Gif-sur-Yvette, France}

\author{E.~Calore}\affiliation{Dipartimento di Fisica e Scienze della Terra, Universit\`a di Ferrara e INFN, Sezione di Ferrara, I-44122  Ferrara, Italy}

\author{A.~Cruz}\affiliation{Departamento  de F\'\i{}sica Te\'orica, Universidad de Zaragoza, 50009 Zaragoza, Spain}\affiliation{Instituto de Biocomputaci\'on y F\'{\i}sica de Sistemas Complejos (BIFI), 50018 Zaragoza, Spain}

\author{L.A.~Fernandez}\affiliation{Departamento  de F\'\i{}sica Te\'orica I, Universidad Complutense, 28040 Madrid, Spain}\affiliation{Instituto de Biocomputaci\'on y F\'{\i}sica de Sistemas Complejos (BIFI), 50018 Zaragoza, Spain}

\author{J.M.~Gil-Narvion}\affiliation{Instituto de Biocomputaci\'on y F\'{\i}sica de Sistemas Complejos (BIFI), 50018 Zaragoza, Spain}

\author{A.~Gordillo-Guerrero}\affiliation{Departamento de  Ingenier\'{\i}a El\'ectrica, Electr\'onica y Autom\'atica, U. de Extremadura, 10071, C\'aceres, Spain}\affiliation{Instituto de Biocomputaci\'on y F\'{\i}sica de Sistemas Complejos (BIFI), 50018 Zaragoza, Spain}

\author{D.~I\~niguez}\affiliation{Instituto de Biocomputaci\'on y F\'{\i}sica de Sistemas Complejos (BIFI), 50018 Zaragoza, Spain}\affiliation{Fundaci\'on ARAID, Diputaci\'on General de Arag\'on, Zaragoza, Spain}

\author{A.~Maiorano}\affiliation{Dipartimento di Fisica, Sapienza
  Universit\`a di Roma, 
  I-00185 Rome, Italy}\affiliation{Instituto de Biocomputaci\'on y F\'{\i}sica de Sistemas Complejos (BIFI), 50018 Zaragoza, Spain}

\author{E.~Marinari}\affiliation{Dipartimento di Fisica, Sapienza
  Universit\`a di Roma, INFN, Sezione di Roma 1, and CNR-Nanotec,
  I-00185 Rome, Italy}

\author{V.~Martin-Mayor}\affiliation{Departamento  de F\'\i{}sica Te\'orica I, Universidad Complutense, 28040 Madrid, Spain}\affiliation{Instituto de Biocomputaci\'on y F\'{\i}sica de Sistemas Complejos (BIFI), 50018 Zaragoza, Spain}

\author{J.~Monforte-Garcia}\affiliation{Instituto de Biocomputaci\'on y F\'{\i}sica de Sistemas Complejos (BIFI), 50018 Zaragoza, Spain}

\author{A.~Mu\~noz-Sudupe}\affiliation{Departamento  de F\'\i{}sica Te\'orica I, Universidad Complutense, 28040 Madrid, Spain}\affiliation{Instituto de Biocomputaci\'on y F\'{\i}sica de Sistemas Complejos (BIFI), 50018 Zaragoza, Spain}

\author{D.~Navarro}\affiliation{Departamento de Ingenier\'{\i}a, Electr\'onica y Comunicaciones and I3A, U. de Zaragoza, 50018 Zaragoza, Spain}

\author{G.~Parisi}\affiliation{Dipartimento di Fisica, Sapienza
  Universit\`a di Roma, INFN, Sezione di Roma 1, and CNR-Nanotec,
  I-00185 Rome, Italy}

\author{S.~Perez-Gaviro}\affiliation{Centro Universitario de la Defensa, Carretera de Huesca s/n, 50090 Zaragoza, Spain}\affiliation{Instituto de Biocomputaci\'on y F\'{\i}sica de Sistemas Complejos (BIFI), 50018 Zaragoza, Spain}

\author{F.~Ricci-Tersenghi}\affiliation{Dipartimento di Fisica, Sapienza
  Universit\`a di Roma, INFN, Sezione di Roma 1, and CNR-Nanotec,
  I-00185 Rome, Italy}

\author{J.J.~Ruiz-Lorenzo}\affiliation{Departamento de F\'{\i}sica and Instituto de Computaci\'on Cient\'{\i}fica Avanzada (ICCAEx), Universidad de Extremadura, 06071 Badajoz, Spain}\affiliation{Instituto de Biocomputaci\'on y F\'{\i}sica de Sistemas Complejos (BIFI), 50018 Zaragoza, Spain}

\author{S.F.~Schifano}\affiliation{Dipartimento di Matematica e Informatica, Universit\`a di Ferrara e INFN, Sezione di Ferrara, I-44122 Ferrara, Italy}

\author{B.~Seoane}\affiliation{Laboratoire de Physique Th\'eorique, \'Ecole Normale Sup\'erieure \& Universit\'e de Recherche Paris Sciences et Lettres, Pierre et Marie Curie \& Sorbonne Universit\'es, UMR 8549 CNRS, 75005 Paris, France}\affiliation{Instituto de Biocomputaci\'on y F\'{\i}sica de Sistemas Complejos (BIFI), 50018 Zaragoza, Spain}

\author{A.~Tarancon}\affiliation{Departamento  de F\'\i{}sica Te\'orica, Universidad de Zaragoza, 50009 Zaragoza, Spain}\affiliation{Instituto de Biocomputaci\'on y F\'{\i}sica de Sistemas Complejos (BIFI), 50018 Zaragoza, Spain}

\author{R.~Tripiccione}\affiliation{Dipartimento di Fisica e Scienze della Terra, Universit\`a di Ferrara e INFN, Sezione di Ferrara, I-44122 Ferrara, Italy}

\author{D.~Yllanes}\affiliation{Department of Physics and Soft Matter Program, Syracuse University, Syracuse, NY, 13244}\affiliation{Instituto de Biocomputaci\'on y F\'{\i}sica de Sistemas Complejos (BIFI), 50018 Zaragoza, Spain}

\collaboration{Janus Collaboration}

\date{\today}

\begin{abstract}
  We first reproduce on the Janus and Janus II computers a milestone
  experiment that measures the spin-glass coherence length through the
  lowering of free-energy barriers induced by the Zeeman effect.
  Secondly we determine the scaling behavior that allows a
  quantitative analysis of a new experiment reported in the companion
  Letter
  \href{https://doi.org/10.1103/PhysRevLett.118.157203}{[S. Guchhait
    and R. Orbach, Phys. Rev. Lett.  {\bf 118}, 157203 (2017)].}  The
  value of the coherence length estimated through the analysis of
  microscopic correlation functions turns out to be quantitatively
  consistent with its measurement through macroscopic response
  functions. Further, non-linear susceptibilities, recently measured
  in glass-forming liquids, scale as powers of the same microscopic
  length.
\end{abstract}

\maketitle 

\paragraph{Introduction.}
It has long been suspected that the exceedingly slow dynamics that
disordered and glassy systems (spin glasses, super-spin glasses,
colloids, polymers, etc.) exhibit upon cooling is due to the
increasing size of the cooperative regions~\cite{adam:65}, which one
would like to describe in terms of a correlation length $\xi$. The
standard way of accessing $\xi$ is measuring the structure factor in a
neutron-scattering experiment. Unfortunately, this approach is
unsuitable for experiments on glassy systems, because their structure
factors show no trace of a growing length scale.

Yet, for example for spin-glass systems, the replica method 
provides a ``microscopic'' approach to obtain the correlation functions
of the overlap
field~\cite{rieger:93,marinari:96,kisker:96,marinari:00b,
  berthier:02,berthier:04,jimenez:05,jaubert:07,janus:08b,janus:09b,liu:14,fernandez:15,lulli:15,manssen:15},
which decay with a correlation length $\xim$. Unfortunately, these
correlation functions are only easy to access through numerical
simulations, since computing replicas requires  direct access to the
microscopic configurations.

In spite of the above difficulties it has been possible to develop
effective techniques to measure $\xi$ in real experiments.  The
state-of-the-art techniques are based on non-linear responses to
external perturbations. Very often these measurements are carried out
in a non-equilibrium regime.  If the temperature is low enough, $\xi$
grows sluggishly but also indefinitely (unless the sample has a film
geometry~\cite{guchhait:14,guchhait:15a}).  For spin glasses and
super-spin glasses, the magnetic response to an external magnetic
field is accurately measured with a SQUID. A delicate analysis of this
response yields a ``macroscopic''
correlation length, which we denote by $\xiM$, as a
function of time.  In the case of glass-forming liquids, one can study
the dielectric polarizability.

Here we implement numerically, for the first time, on the Ising spin
glass, the seminal experimental protocol introduced
in~\cite{joh:99}, which is now a crucial protocol for spin glass
experiments~\cite{guchhait:14,nakamae:12}. Thanks to our dedicated
computers Janus \cite{janus:08} and Janus II \cite{janus:14}, the
system size and the time scales reached in our simulation allow us to
assert the mutual consistency of the correlation lengths obtained
through macroscopic response, $\xiM$, and the length scale $\xim$
derived from the direct measurement of the overlap correlation
function.

Our analysis unveils a scaling law describing how the magnetic
response depends both on the applied magnetic field $H$ and on the
size $\xim$ of the magnetic domains. Remarkably, this scaling law is
already very useful in the analysis of the experiment by Guchhait and
Orbach described in the companion Letter~\cite{guchhait:17}. 

 The reader is probably aware of the long ongoing controversy about the
nature of the spin-glass phase. The Replica Symmetry Breaking
theory~\cite{marinari:00} predicts a spin-glass transition in a
field~\cite{dealmeida:78}, while the droplet model predicts that the
magnetic field (no matter how small) avoids the
transition~\cite{mcmillan:84,bray:87,fisher:86,fisher:88}. In
particular, the dynamics of a spin-glass in a field has been analyzed
within the context of the droplet model~\cite{takayama:04}. However,
it has been difficult for experiments to distinguish both
theories~\cite{lefloch:92,jonason:98,joh:99,jonason:00,bouchaud:01},
because the two predict a barrier-height that depends on the length
scale $\xim$. Fortunately, our analysis completely avoids this controversy.

Finally, we link our results to the physics of
glass-forming liquids through a
study of the non-linear susceptibilities $\chi_3$ (see below). To
date it has not yet been possible to reproduce the delicate
experimental protocol of Ref.~\cite{joh:99} for supercooled liquids or
glasses. However, $\chi_3$~\cite{berthier:05} (and also
$\chi_5$~\cite{albert:16}) can be measured and do grow. We find that
in our spin-glass simulation $\chi_3$ has a well-defined scaling form
as a power of $\xim$.

\paragraph{Model and protocol.}
We study the Edwards-Anderson model in a three-dimensional, $D=3$
cubic lattice of linear size
$L$, with periodic boundary conditions. Our $N=L^D$ Ising spins,
$\sigma_{\boldsymbol{x}}=\pm 1$, interact with their lattice nearest
neighbors through the Hamiltonian
\begin{equation}\label{eq:EA-H}
{\cal H}=-\sum_{\langle \boldsymbol{x}, \boldsymbol{y}\rangle } J_{\boldsymbol{x},\boldsymbol{y}} \sigma_{\boldsymbol{x}}\, \sigma_{\boldsymbol{y}}-H\sum_{\boldsymbol{x}}\sigma_{\boldsymbol{x}}\,.
\end{equation}
The couplings $J_{\boldsymbol{x},\boldsymbol{y}}$ take the values $\pm 1$ with
$50\%$ probability. In the absence of a magnetic field, $H=0$, this
model undergoes a spin-glass transition at the critical temperature
$T_\mathrm{c}=1.102(3)$~\cite{janus:13}. The value of the
dimensionless magnetic field
$H$ used in the numerical simulation can be matched to the physical one.  For
the Ising spin glass Fe$_{0.5}$Mn$_{0.5}$TiO$_3$ we find
$H_\mathrm{experimental}\approx 50$ kG $\times H$~\footnote{
  At the critical temperature, we matched the value of 
  \unexpanded{$\left.(\langle m\rangle/H)\right|_{H}/\left. (\langle m\rangle/H)\right|_{H\to 0}$}, 
  computed numerically \cite{janus:14c}, with the same quantity measured
  experimentally in Fe$_{0.5}$Mn$_{0.5}$TiO$_3$~\cite{aruga_katori:94}
  (\unexpanded{$\langle m\rangle$} is the equilibrium value of the
  magnetization density).}. This matching
is likely to be strongly dependent on the material under
consideration.

We describe succinctly our simulation protocol (for details
see the analysis of the aging linear response in~\cite{janus:16}).
We consider a large system (with $L=80$ or $160$, large enough to
avoid relevant finite-size effects). The initial random spin
configuration is placed instantaneously at the working temperature
$T=0.7\approx 0.64 T_\mathrm{c}$ and left to relax for a time $\tw$,
with $H=0$. At time $\tw$, the magnetic field is turned on and we
start recording the magnetization density,
$m=\sum_{\boldsymbol{x}}\sigma_{\boldsymbol{x}}/N$. We write
$m(t+\tw,\tw;H)$ to emphasize that the system is perennially out of
equilibrium (and, hence, $\tw$-dependent).  In the following
the symmetry under the
inversion of the magnetic field, $m(t+\tw,\tw;H)=-m(t+\tw,\tw;-H)$,
will be crucial.
\begin{figure}[tb]
\includegraphics[width=1.0\columnwidth]{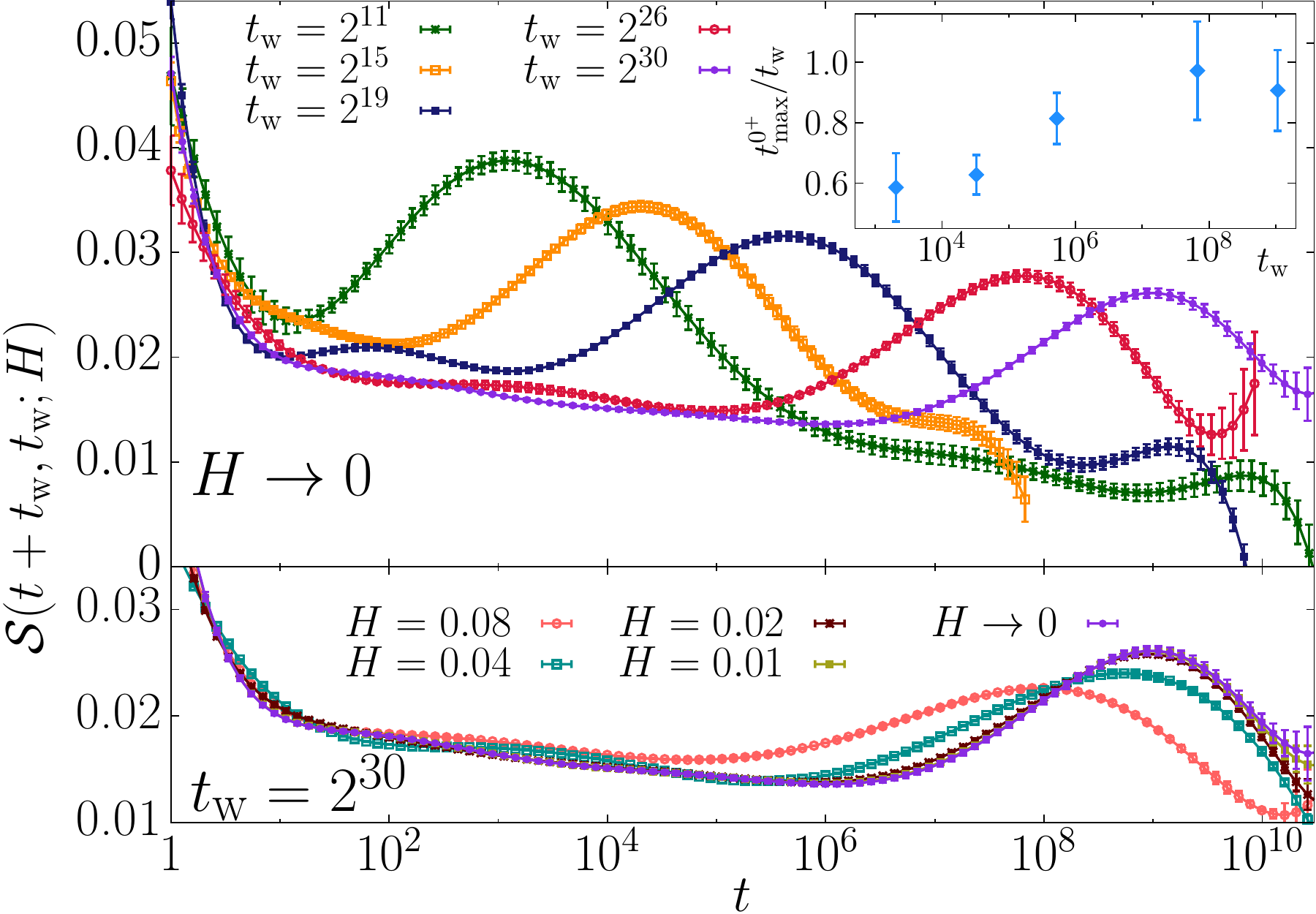}
\caption{The function $\mathcal{S}(t+\tw,\tw;H)$,
  Eq.~\eqref{eq:S-def} versus the time $t$ elapsed after
  switching on the external magnetic field $H$. In the {\bf top} panel
  we show the $H\to 0$ extrapolation for several waiting times $\tw$
  (one unit of computer time roughly corresponds to one
  picosecond of physical time~\cite{mydosh:93}). {\bf Bottom:}
  $\mathcal{S}(t+\tw,\tw;H)$ as a function of $t$ for
  our largest waiting time $\tw=2^{30}$ and for different values of
  $H$.
  {\bf Inset:} The peak position ($H\to 0$), in units of $\tw$, depends on $\tw$
  only for $\tw<10^6$.}
\label{fig:fig1}
\end{figure}

\paragraph{Scaling.} 
As the system relaxes at the working temperature for a time $\tw$, the
size of the glassy domains grows. The overlap correlation function
$C_4(r,\tw)$~\footnote{ $C_4(r,\tw)=\overline{[ q_i(\tw) q_{i+r}(\tw)]}$ where
  $q_i(\tw)$ is the local overlap defined as $q_i=\sigma_i \tau_i$,
  where $\sigma_i$ and $\tau_i$ are two real replicas which evolve
  with the same (random) couplings.  $[(\cdots)]$ denotes the average
  over initial spin configurations and $\overline{(\cdots)}$ is the
  average over the random couplings.} decays with the distance
$r$ as $C_4(r,\tw)=f_\mathrm{c}(r/\xim(\tw))r^{-\theta}$
\cite{marinari:96,janus:08b,janus:09b}.  The cut-off function
$f_\mathrm{c}(x)$ decays faster than exponentially at large $x$. The
exponent $\theta=0.38(2)$~\cite{janus:09b,janus:10} will be crucial in
our analysis. The microscopic coherence length grows with time as
$\xim(\tw)\propto \tw^{1/z(T)}$, with
$z(T=0.7)=11.64(15)$~\cite{janus:09b}.

In equilibrium conditions and for large $\xim$, there is a well
developed scaling theory for the magnetic response to an external
field, see, e.g.,~\cite{parisi:88,amit:05}. However, dynamic
scaling~\cite{ozeki:07} suggests borrowing the equilibrium formulae,
and replacing the equilibrium $\xim$ by the aging  $\xim(t+\tw)$ (as
obtained at $H=0$). This bold approach has been successfully tested
for spin glasses close to $T_\mathrm{c}$~\cite{lulli:15,fernandez:15}
(and, to a small extent, also for glass-forming
liquids~\cite{albert:16}), thanks to the relation
\begin{equation}\label{eq:scaling-borrowed}
m(t+\tw,\tw;H)= \xim^{y_h-D} \mathcal{F}\Big(H [\xim(t+\tw)]^{y_h},{\cal R}_{t,\tw}\Big)\,,
\end{equation}
where $y_h$ is a scaling dimension that we will now determine,  ${\cal
  R}_{t,\tw}\equiv\xim(t+\tw)/\xim(\tw)$, and the scaling function
$\mathcal{F}(x,{\cal R})$ is odd on its first argument for symmetry
reasons. As we will show below, see Fig.~\ref{fig:fig1}--inset, we shall
be interested in the regime $t\approx \tw$ where the approximation
${\cal R}_{t,\tw}\approx 1$ is safe~\cite{janus:16}. Therefore,
$\xim(\tw)$ will be the relevant length scale from now on.

The (generalized) susceptibilities $\chi_1,\chi_3,\chi_5,\ldots$ are
defined from the Taylor expansion
\begin{equation}\label{eq:suscept-defined}
m(H)=\chi_1 H+ \frac{\chi_3}{3!} H^3+ \frac{\chi_5}{5!} H^5+\mathcal{O}(H^7)\,,
\end{equation}
where  we omitted the $t$ and $\tw$ dependencies of $m$ and of
the susceptibilities
to simplify our notation. Matching Eqs. \eqref{eq:scaling-borrowed}
and \eqref{eq:suscept-defined}, we find the scaling behavior
$\chi_{2n-1}\propto [\xi(\tw)]^{2y_hn-D}$.
At least in equilibrium, $\chi_3$ is connected to the
space-integral
of the microscopic correlation function $C_4(r,\tw)$~\cite{binder:86}.
We thus conclude that
\begin{equation}\label{eq:matching-yh}
2 y_h= D-\frac{\theta}{2}\,.
\end{equation}
Taking $\theta$ from~\cite{janus:09b,janus:10}, we find $2
y_h=2.81(1)$.  Although $2 y_h$ is sometimes referred to as the
fractal dimension of the glassy
domains~\cite{berthier:02,albert:16,stevenson:06,nakamae:12}, we
regard it as just a scaling dimension~\footnote{Indeed, a satisfactory
  geometric construction would build spin clusters with a number of
  spins scaling as the relevant susceptibility, i.e.,
  $[\xi(\tw)]^{D-\theta}$ in our case. For ferromagnetic systems,
  these domains are the Fortuin-Kasteleyn (FK)
  clusters~\cite{fortuin:72}. Unfortunately, it is still unknown how
  to satisfactorily generalize FK clusters to glassy systems (for
  instance, the straightforward generalization of FK clusters to spin
  glasses are space-filling ---i.e., percolating--- well above
  $T_\mathrm{c}$~\cite{coniglio:91}). Yet, see
  Refs.~\cite{houdayer:01,houdayer:04,jorg:05,machta:08,zhu:15} for
  some recent work.} (the droplet model prediction is $2y_h=D$).

\paragraph{Simulating the experiment.} The main quantity used
in the experiment of \cite{joh:99} is
\begin{equation}\label{eq:S-def}
  \mathcal{S}(t+\tw,\tw;H)=\frac{\partial}{\partial\log t}
  \left[\frac{m(t+\tw,\tw;H)}{H}\right]\,.
\end{equation}
This quantity, shown in Fig.~\ref{fig:fig1}, has a local maximum at
time $t^{(H)}_\mathrm{max}$. The time scale $t^{(H)}_\mathrm{max}$ was
interpreted by Joh et al. as representative of the free-energy
barriers $\varDelta(\tw;H)$ that are relevant at time $\tw$:
$t^{(H)}_\mathrm{max}\propto
\mathrm{exp}[\varDelta/k_\mathrm{B}T]$~\cite{joh:99}  (see also the
numerical computation in Ref.~\cite{takayama:04}).

$\mathcal{S}(t+\tw,\tw;H)$ depends on two time scales, $t$ and $\tw$, as it is
typical of aging systems~\cite{vincent:97}.  However, we want to use
$\mathcal{S}$ to extract information from the single-time $\xiM(\tw)$. The
paradox is solved in the inset to Fig.~\ref{fig:fig1}, where we show that,
when $\tw$ is large enough,  the ratio $t^{(0^+)}_\mathrm{max}/\tw$ becomes
independent of $\tw$: we are, in these conditions, in the asymptotic regime.
This regime is also reached, at significantly shorter $\tw$, with Gaussian
couplings~\cite{takayama:04}.
\begin{figure}[tb]
\includegraphics[width=1.0\columnwidth]{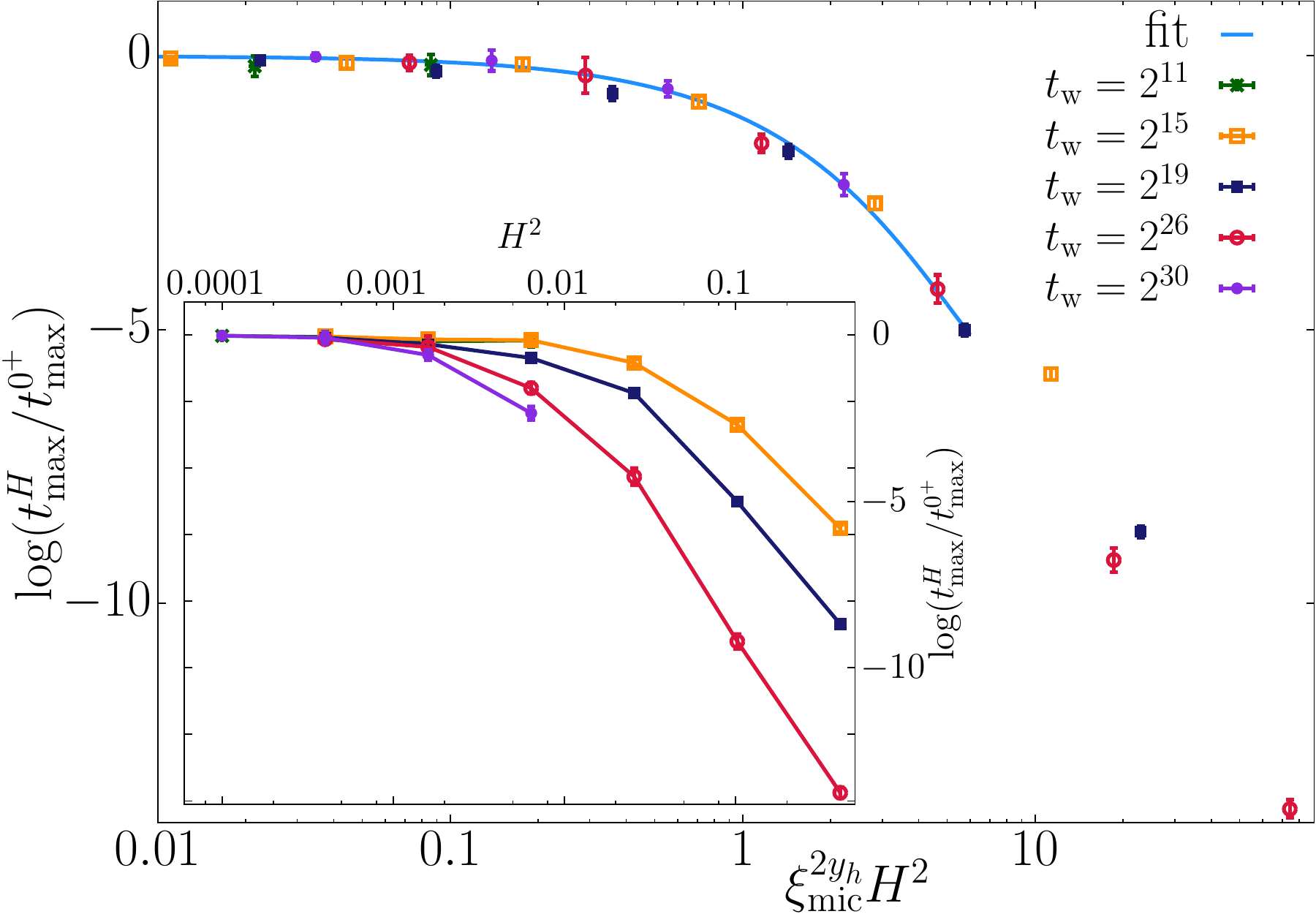}
\caption{The Zeeman energy follows the scaling form suggested in
  Eq.~\eqref{eq:scaling-Zeeman}. We show a fit to
  $F_\mathrm{Zeeman}(x)=c_1 x+c_2 x^2$.  {\bf Inset:} the data of the
  main panel do not collapse when plotted as a function of $H^2$.}
\label{fig:fig2}
\end{figure}

The maximum $t^{(H)}_\mathrm{max}$ decreases upon increasing $H$, see
Fig.~\ref{fig:fig1}--bottom. This reflects the lowering of the
barriers $\varDelta$ due to the Zeeman effect of the (glassy) magnetic
domains~\cite{joh:99}.  From Eq.~\eqref{eq:scaling-borrowed}, and
given the $H\leftrightarrow -H$ symmetry, it is natural to expect 
the Zeeman effect to be described through a smooth scaling function
\begin{equation}\label{eq:scaling-Zeeman}
  \log\frac{t^{(H)}_\mathrm{max}}{t^{0^+}_\mathrm{max}} =
  F_\mathrm{Zeeman}(x)\,,\ x=H^2[\xim(\tw)]^{D-\frac{\theta}{2}}\,,
\end{equation} 
where $t^{0^+}_\mathrm{max}$ is the extrapolation to $H=0$ of
$t^{(H)}_\mathrm{max}$.  As Fig.~\ref{fig:fig2} shows, this scaling
holds for values of the scaling variable as large as $x\approx 6$: we
have a very good scaling for close to three orders of magnitude.
Up to that value, we find that the scaling function can be
parameterized as $F_\mathrm{Zeeman}(x)=c_1 x + c_2 x^2$. In other
words, for small $H$ we expect the Zeeman energy to be proportional to
$H^2$ with sizable corrections of order $H^4$.
\begin{figure}[t]
\includegraphics[width=1.0\columnwidth]{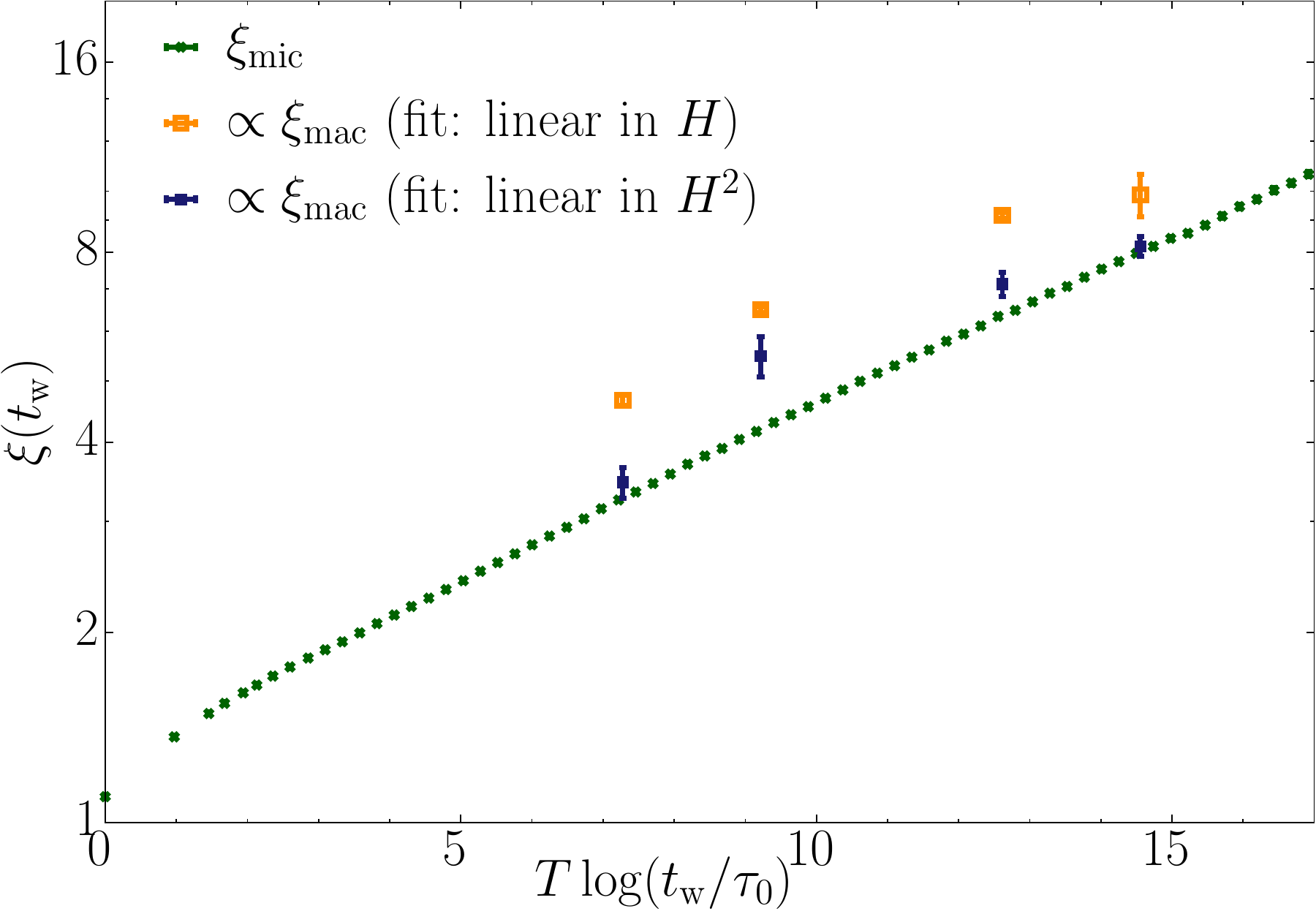}
\caption{The time growth of the correlation length $\xim$, as obtained
  from the microscopic correlation function
  $C_4(r,\tw)$~\cite{janus:08b,janus:09b,janus:16}, is compared to the
  length $\xiM$ obtained from a fit linear in $H^2$, see
  Eq.~\eqref{eq:exp-fit1}. The microscopic time scale $\tau_0=1$
  corresponds to a single lattice sweep in our Monte Carlo simulation
  (see caption to Fig.~\ref{fig:fig1}).  We also show the results
  obtained with Eq~\eqref{eq:exp-fit3}, which are sensible as
  well.  The temperature-dependent scaling variable, $T \log(\tw/\tau_0)$, is
  common in the experimental literature (e.g., see
  Ref. \cite{nakamae:12}).}
\label{fig:fig3}
\end{figure}
To the best of our knowledge, the explicit scaling form in
Eq.~\eqref{eq:scaling-Zeeman} has never  been used in the analysis of
experimental data. Yet the authors of the
original experiment~\cite{joh:99} fitted their data at fixed $\tw$ to
\begin{equation}\label{eq:exp-fit1}
\log\frac{t^{(H)}_\mathrm{max}}{t^{0^+}_\mathrm{max}}=A N_f(\tw) H^2\,,
\end{equation}
where $A$ is a $\tw$-independent constant.
$N_f(\tw)$ was interpreted as the number of spins in a correlated
domain, and hence
\begin{equation}\label{eq:exp-fit2}
 \xiM(\tw)=[N_f(\tw)]^{1/D}\,.
\end{equation}
Eqs. \eqref{eq:exp-fit1} and \eqref{eq:exp-fit2} can be seen
as the first-order expansion of Eq.~\eqref{eq:scaling-Zeeman}. In fact, the
smallness of exponent $\theta$ implies that the small correction
$[\xiM(\tw)]^{\theta/2}$ can easily go unobserved. 

Fig.~\ref{fig:fig3} shows
  $\xiM(\tw)=[N_f(\tw)]^{1/(D-\theta/2)}$ [we obtained $N_f(\tw)$ from
    the fit to Eq.~\eqref{eq:exp-fit1}]. Since different determinations of the
  correlation length should coincide only up to a multiplicative
  constant of order one, we have not fitted for $A$, choosing instead
  $A=1$. It is clear that $\xiM(\tw)$ and $\xim(\tw)$ have the same
  behavior.

Finally, let us remark that in Ref.~\cite{bert:04} it was suggested
that Ising spin glasses should have a Zeeman energy of order $H$. On
theoretical grounds, this is not possible for protocols respecting the
symmetry $H\leftrightarrow -H$. However, we found that for $1<x<4$
a best
fit to the form $F_\mathrm{Zeeman}(x)=d_1 +d_2\sqrt{x}$ gives an
acceptable value of $\chi^2$, but one gets that $d_1\neq 0$,
that implies an unphysical value for 
the  $H\to 0$ extrapolation.
Only a careful control of the limit of vanishing field (see the companion
Letter by Guchhait and Orbach~\cite{guchhait:17}), reveals that the true behavior for
small $H$ is proportional to $H^2$. In practice, the
transient behavior of $F_\mathrm{Zeeman}(x)$ implies that one could
fit the data to the form
\begin{equation}\label{eq:exp-fit3}
\log\frac{t^{(H)}_\mathrm{max}}{t^{0^+}_\mathrm{max}}=A'\sqrt{N_f(\tw) H^2}\,,
\end{equation}
and then extract  $\xiM(\tw)=[N_f(\tw)]^{1/(D-\theta/2)}$
  (again, $A'=1$).  Although Eq.~\eqref{eq:exp-fit3} is incorrect for
small values of $H$, the scaling law Eq.~\eqref{eq:scaling-Zeeman}
implies that one will still obtain a reasonable determination of
$\xiM$, as we indeed find (see Fig.~\ref{fig:fig3}, where we also show
$\xiM(\tw)$ obtained from this approach).

\paragraph{Non-linear susceptibilities.}
 At variance with spin glasses~\cite{joh:99}, the detection of a large
correlation length accompanying the glass transition is still an open
problem for supercooled liquids~\cite{karmakar:14}. It is now clear
that linear responses are not up to the
task~\cite{berthier:05,biroli:08}, so higher-order non-linear responses
are currently under
investigation~\cite{berthier:05,albert:16,brun:12}.
However, even in the more familiar context of spin glasses the
connection between $\chi_3(t+\tw;\tw)$ and $\xim(\tw)$ needs to be
clarified. 

\begin{figure}[!tbh]
\includegraphics[width=1.0\columnwidth]{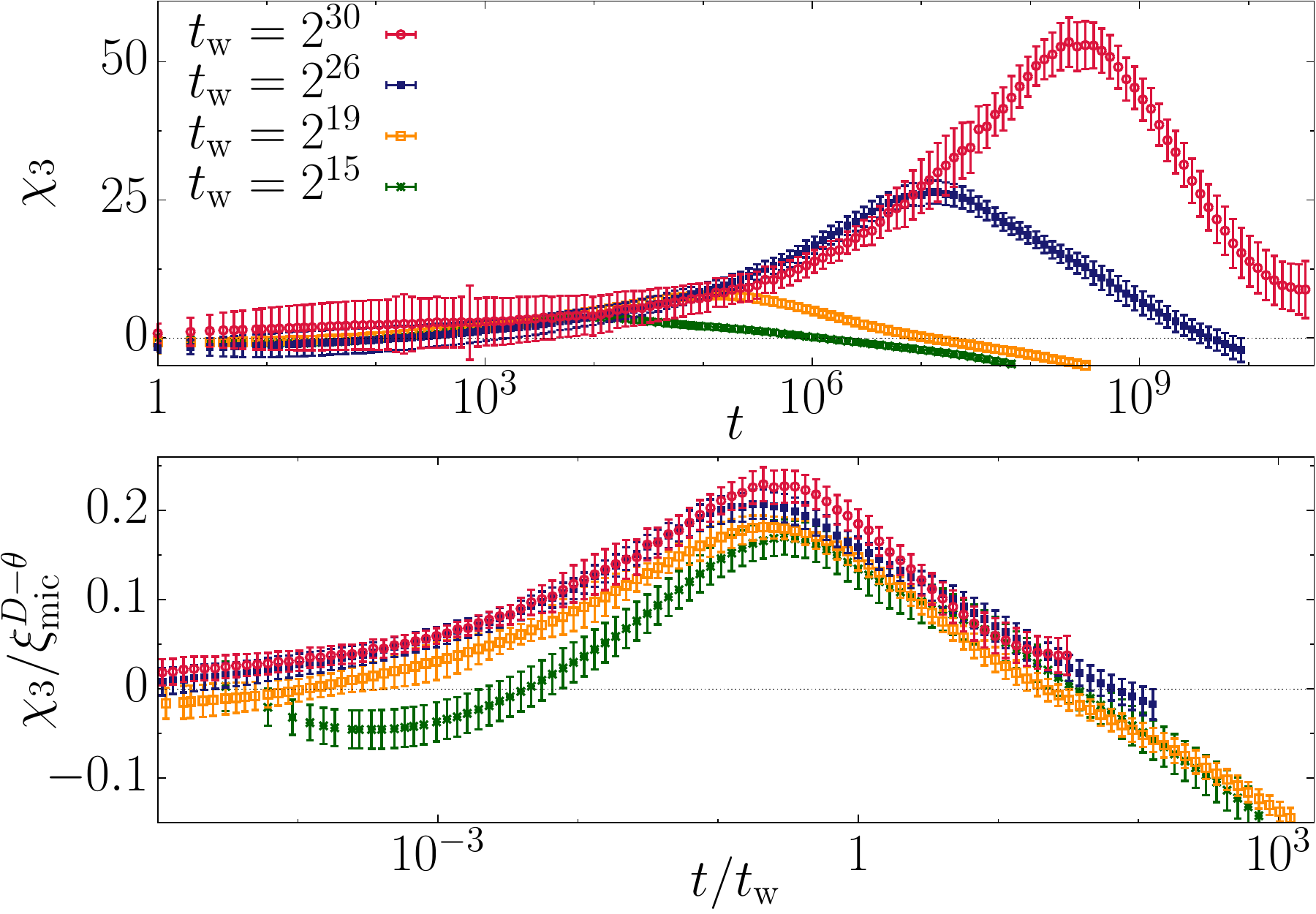}
\caption{The non-linear susceptibility $\chi_3$ is shown as a function
  of $t$ for several values of $\tw$ ({\bf top}), as
    obtained from Eq.~\eqref{eq:suscept-defined}.  The difficulty lies
    in balancing systematic errors (numerical data obtained with high
    fields underestimate $\chi_3$) with statistical errors (which are
    larger for small values of $H$). Our compromise, shown here, tries
    to obtain statistical and systematic errors of comparable size.
  In the {\bf bottom} panel we show the $G(t/\tw)$ scaling
  function~\eqref{eq:chi3-scaling}. Note that scaling corrections are
  visible only for the smallest waiting times (and, even in those
  cases, they only appear for small $t/\tw$).}
\label{fig:fig4}
\end{figure}

To make some progress, we extract generalized susceptibilities such as
$\chi_3$ through Eq.~\eqref{eq:suscept-defined}.
Fig.~\ref{fig:fig4}--top shows that $\chi_3(t+\tw,\tw)$ has a
$\tw$-independent regime for $t\ll\tw$ (the time-translational
invariant regime~\cite{vincent:97}, see also~\cite{biroli:16}). Yet,
it displays a peak as a function of $t$, whose position and height are
strongly $\tw$-dependent. In fact, we empirically find (see
Fig.~\ref{fig:fig4}-bottom) the following scaling behavior for large
enough values of $t$ and $\tw$
\begin{equation}\label{eq:chi3-scaling} \chi_3(t+\tw,\tw)=
[\xim(\tw)]^{D-\theta} G(t/\tw)\,.  \end{equation} The prefactor
$[\xim(\tw)]^{D-\theta}$ follows from Eqs. \eqref{eq:scaling-borrowed},
\eqref{eq:suscept-defined} and \eqref{eq:matching-yh}.  Deriving the details
of the function $G(t/\tw)$ will require further work.

\paragraph{Conclusions.}  Using the dedicated computers Janus and Janus II, 
we have studied the aging magnetic response of an Ising spin-glass to
an applied field. In this way, we have simulated a milestone
experiment~\cite{joh:99}, and we have shown that the glassy correlation
length extracted from this macroscopic response is numerically
consistent with its microscopic determination from overlap correlation
functions. Furthermore, we have unveiled scaling laws that relate the
magnetic response to the applied field and the correlation length. We
expect that this scaling analysis will be useful in future
experiments on film geometry. Our scaling analysis has  been
relevant for the study of the experiment reported in the
companion Letter~\cite{guchhait:17}. The agreement with experiments is even more
impressive when one notices that we are comparing numerical time
scales of the order of the millisecond to experimental time scales of
the order of the hour: this looks like a very nice piece of evidence for
invariance in time scales.

Although the delicate experimental study of
Ref.~\cite{joh:99} has not yet been carried out for glass-forming
liquids, the (dielectric polarizability analogue of) the non-linear
susceptibilities are measured in current
experiments~\cite{berthier:05,albert:16}. We have shown that
these susceptibilities scale as powers of the
microscopically-determined correlation lengths.

\begin{acknowledgments}
\paragraph{Acknowledgments.}
We warmly thank Ray Orbach and Samaresh Guchhait for sharing with
us their data prior to publication~\cite{guchhait:17}, and for a most fruitful
exchange of ideas.

We thank the staff of BIFI supercomputing center for their assistance.
We thank  M. Pivanti for his  contribution to the early  stages of the
development of the  Janus II computer. We also  thank Link Engineering
(Bologna,  Italy) for  their precious  role in  the technical  aspects
related to  the construction of Janus  II. We thank EU,  Government of
Spain and  Government of Aragon  for the financial support  (FEDER) of
Janus II development.  This work was partially supported by Ministerio
de  Econom\'ia, Industria  y Competitividad  (MINECO) (Spain)  through
Grants     No.     FIS2012-35719-C02,    No.     FIS2013-     42840-P,
No. FIS2015-65078-C2, No. FIS2016-76359-P, and No. TEC2016-78358-R, by
the Junta de Extremadura (Spain) through Grant No. GRU10158 (partially
funded by FEDER) and by  the DGA-FSE (Diputaci\'on General de Arag\'on
-- Fondo Social  Europeo). This project has received  funding from the
European Union's  Horizon 2020  research and innovation  program under
the  Marie Sk{\l}odowska--Curie  Grant  No. 654971.  This project  has
received funding  from the European  Research Council (ERC)  under the
European Union's  Horizon 2020 research and  innovation program (Grant
No. 694925).  D. Y. acknowledges  support by Grant  No. NSF-DMR-305184
and     by     the     Soft     Matter     Program     at     Syracuse
University. M.  B. J.  acknowledges financial  support from  ERC Grant
No. NPRGGLASS.

\end{acknowledgments}

\bibliographystyle{apsrev4-1}

\bibliography{/homenfs/rg/biblio}

\end{document}